\documentclass[twocolumn,amsmath,amssymb]{revtex4}
\usepackage{graphicx}% Include figure files

\usepackage{color}
\usepackage{ hyperref}
\hypersetup{%
        colorlinks,
        linkcolor={blue},
        citecolor={red},
        urlcolor={blue}
                      }%--------------------------------------------------------------
\def\be{\begin{equation}}
\def\ee{\end{equation}}
\def\bee{\begin{eqnarray}}
\def\eee{\end{eqnarray}}
\begin{document}
\author{Giuseppina Modestino}
\email{modestino@lnf.infn.it}
\affiliation{%
INFN, Laboratori Nazionali di Frascati, via Enrico Fermi 40, I-00044, Frascati (Roma) Italy\\
}%
\date{\today}% It is always \today, today,
\title{ Explanation of the Special Theory of Relativity by Analytical Geometry and Reformulation of the Inverse-Square-Law}

\date{\today }

\begin{abstract}
The space-time length  $R$ between  a moving source and the observation point is calculated in order to substitute with it the spatial distance $D$, normally used in the Newton's law of gravitation, as well as in any $inverse$-$square$-$law$. Fundamentally, three space-time amounts describe dynamics. The relationship between position and field intensity is analytic, 
estimable in euclidean space, and considering a linear reference system for the time parameter.  The formulation shows compatibility  with fundamental rules  of classical mechanics, highlighting also hitherto unknown properties, as a perfect analogy between morphological and physical parameters, such as the complete correspondence between the eccentricity and the momentum in the orbital motion. Moreover, the procedure naturally contains relativistic formulation without introducing any special hypothesis on light speed isotropy, asking so the question about the actual need to introduce the concept of space-time curvature for the correct interpretation of physics phenomena.  
\end{abstract}
\maketitle
\section{Introduction}
\label{intro}
As physical theory, special relativity (SR) \cite{ein1} has been experimentally well tested, and until now  inconsistencies don't result. Principally generated for explain the results Michelson-Morley experiment \cite{mimo}, SR is built on two postulates, one regarding  the invariance of the physical laws with respect to the inertial systems, the other one regarding the constant speed of light observed from any point of view, even if source and observer are relatively moving.  Consequently, for fitting with the real experience, the concept of space-time curvature is introduced. The scientific literature is  full of experimental evidences supporting the relativistic hypotheses (\cite{what} and references within). Nevertheless, SR remains a model-dependent environment, since observations do not have an independent status from the theoretical content in which they appear. At least, observation results are used for replace the Einstein's postulates, anyway with the limitation of inductive method, for instance as done by Robertson \cite{robe}. The present study concerns fundamentally typical relativistic problems as the signal evaluation from a moving source but regardless Einstein's postulates neither other conventional procedures. Synthesized by $R$, the new distance concept will be compared to the classical one represented by $D_t$, and it will be applied to classical mechanics laws, thus proving its effectiveness.
\section{moving signal observation}
\label{example1}
 \subsection{Reference system $S(x,y)$}
 \label{rif}
\begin{figure}
\includegraphics[width=7.5cm]{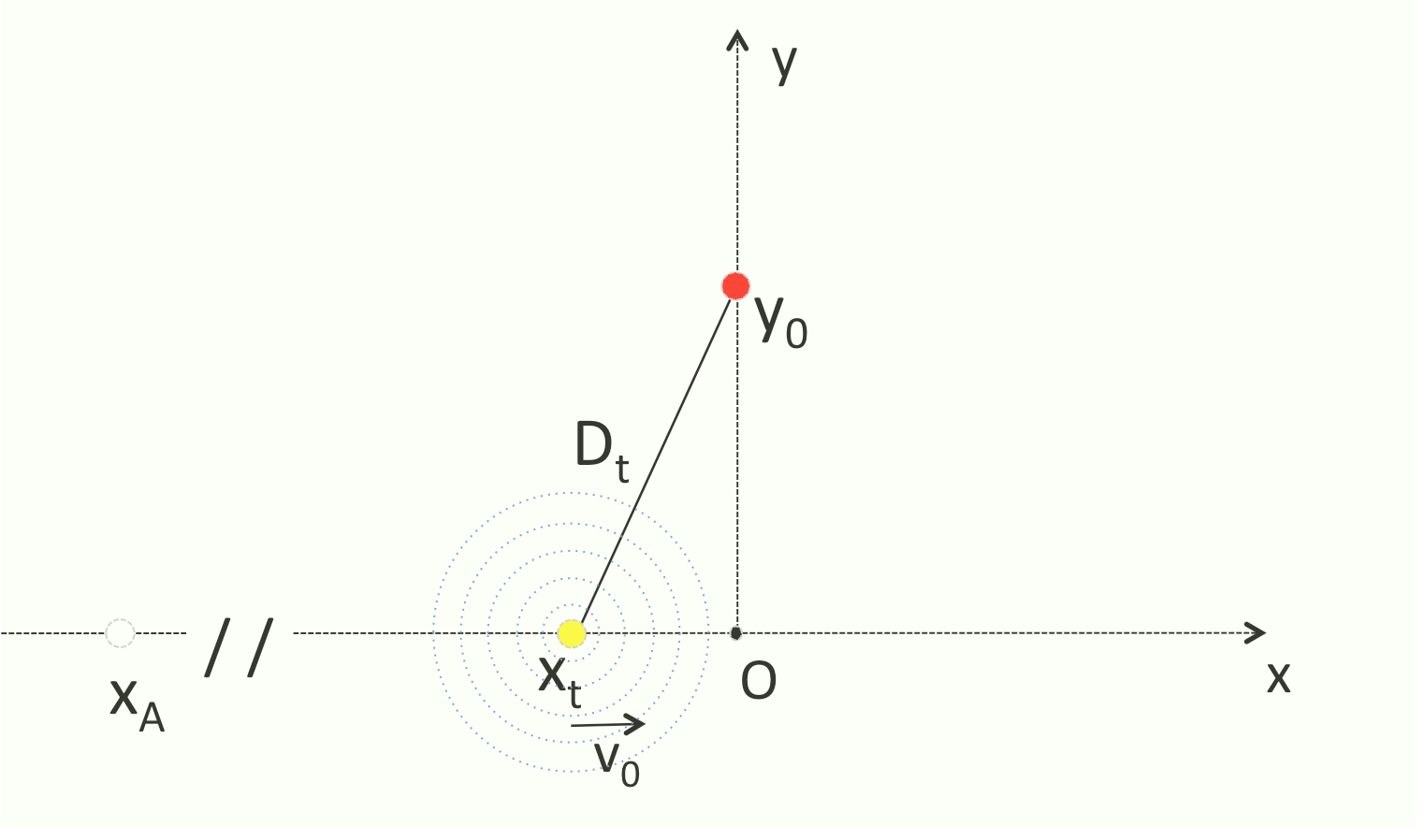}
\caption{Coordinates of a source emitting a radial signal at light speed c, and the observer point, respectively $[x_t,0]$ and $[0,y_0]$. The source is moving at constant velocity $v_0$ . The coordinate $x_A$ represents the source position at time $t=0$ in the absolute reference line $T_A$, or equivalently, at $t=x_A/v_0$ in the translate reference line $T$ (see the next figure).
\label{squa1} }
\end{figure}

Let's consider a  constant power source that emits a signal propagating at radial velocity $c$ \cite{luce}. It is a point source, in such a way no interference phenomenon will occur. The source moves with constant velocity $v$ in a reference system $S(x,y)$ (choosing a two-dimensional space with z = 0 ) in which an observer  is placed at $[0,y_0]$. The motion occurs in a direction parallel to the x-axis, being $[x_t,0]$ the coordinates of the source, and ${\bf v_0}\equiv [v_0,0]$. The distance between source and observer is \be
D_t=\sqrt{x_t^2+y_0^2}.
\label{dtt}
\ee
The framework is illustrated in fig.\ref{squa1}
\subsection{Reference time line $T$}
\begin{figure}
\includegraphics[width=7.5cm]{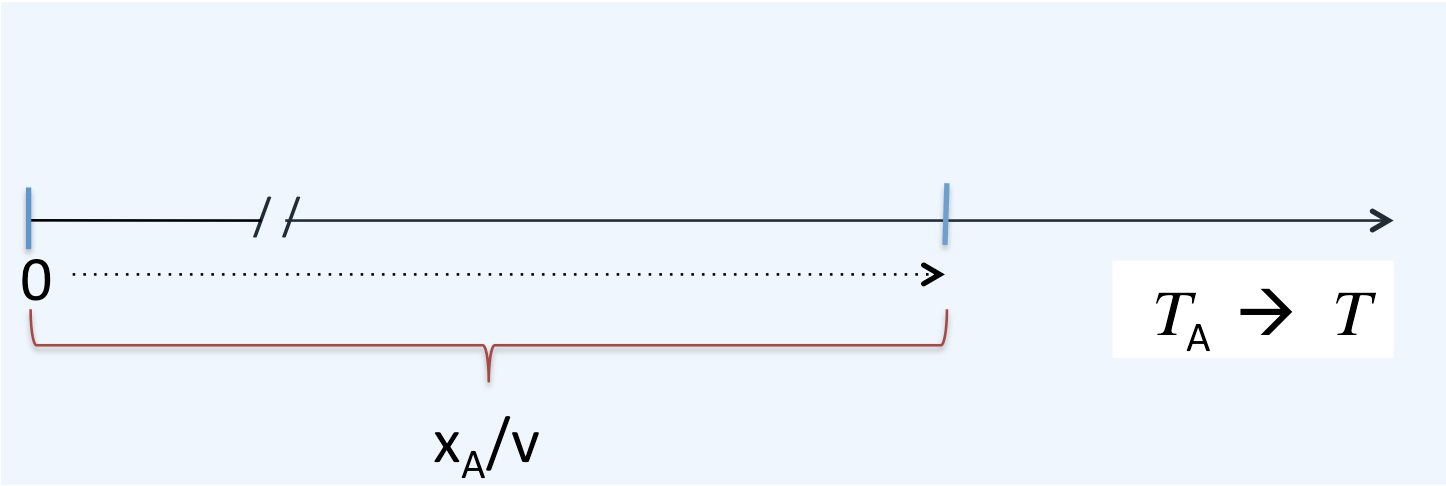}
\caption{Time line on which a reference translation from $T_A$ to $T$ has been performed in such a way the eq.\ref{xt2} holds.
\label{reftime} }
\end{figure}

The position of the source is determined by the relationship 
\be
x_t=x_{A}+v_0t_A~~~~~~~~{\text if} ~~~~t_A~\in T_A
\label{xt1}
\ee
where $x_{A}$ is the source position at time $t=0$, and $T_{A}$ is the absolute reference time line, with origin at $0$. Let's perform a translation on the temporal reference system 
\be
T_A\rightarrow T,
\label{t_arrow}
\ee
 that is 
 adding the constant length to the origin (fig.\ref{reftime})  
 \be
 t=t_A+x_A/v_0,
 \label{t_rel}
 \ee
 such that the temporal law of eq.\ref{xt1} becomes
\be
x_t=v_0t ~~~~~~~~\text{being} ~~~~t~\in T.
\label{xt2}
\ee
In spite of the triviality of the last step, it is important underline that the choice of time reference system is crucial since it must remain unchanged in all developments of the system dynamics.
\subsection{Intensity field and R parameter}
\label{spacetime}
 Due to the emission, an intensity field $E_t$ is recorded at $[0,y_0]$. 
The purpose of the following calculation is to find the relationship between $x_t$ and $E_t$ within the defined space, and respecting the physical rates $v_0$ and $c$.
Assuming worth the $inverse~square~law$, the intensity is
\be
E_t \propto 1/R^2, 
\label{et1}
\ee
where 
\be
R\equiv c\Delta t, ~~~~{\text{ and}}~~~~ \Delta t\equiv t-t_0
\label{rt1}
\ee
 is the time interval during which the signal is born, at $t_0$, and grows at the speed $c$ until the time $t$, when it reaches the observer. 
Having well defined the space-time relationship in the eq.\ref{xt2}, now to establish a correct prediction of $E_t$ is possible as a function of time, or equivalently of the spatial coordinates of the chosen reference system.
When both observer and source are at rest, that is also $v_0=0$, $R$ simply corresponds to space distance $D_t$, then the Coulomb's law applies, i.e. $E_t\sim 1/D_t= const.$ Treating a dynamic case, the correspondence is not satisfied. In that case, it is useful to consider a previous time instant 
\be
t_p\equiv t-D_t/c ~~~{\text with} ~t_p \in T,
\label{tpi}
\ee
and relative source position 
\be
x_p\equiv v_0t_p,
\label{xpi} 
\ee
 such that 
\be
x_t=x_p+\beta_0 D_t=x_p+\beta_0\sqrt{x_t^2+y_0^2}
\label{xpxt}
\ee
and
\be
t-t_p=D_t/c=\frac{\sqrt{v_0^2 t^2+y_0^2}}{c}
\label{xp1}
\ee
with $\beta_0\equiv v_0/c$.

\begin{figure}
\includegraphics[width=1.\linewidth]{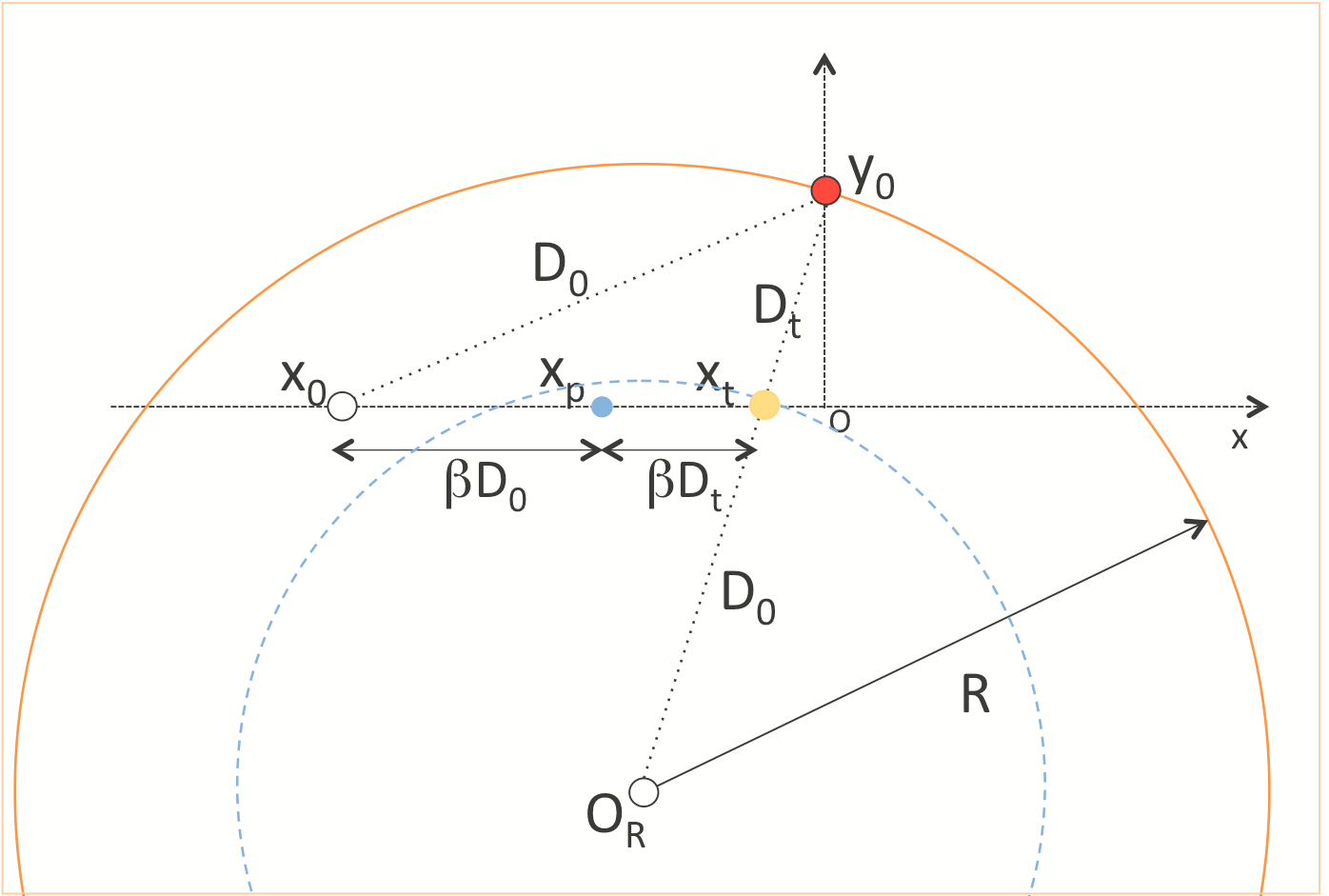}
\caption{
The figure shows all the spatial parameters defined in section \ref{spacetime}. The point $O_R$ is a the virtual origin of the path $R$ covered by the signal, at $c$ speed. 
%The speed of the source is $v_0\approx 0.6 c$, and$y_0=0.3m$. 
\label{spacex} }
\end{figure}

\begin{figure}
\includegraphics[width=1.\linewidth]{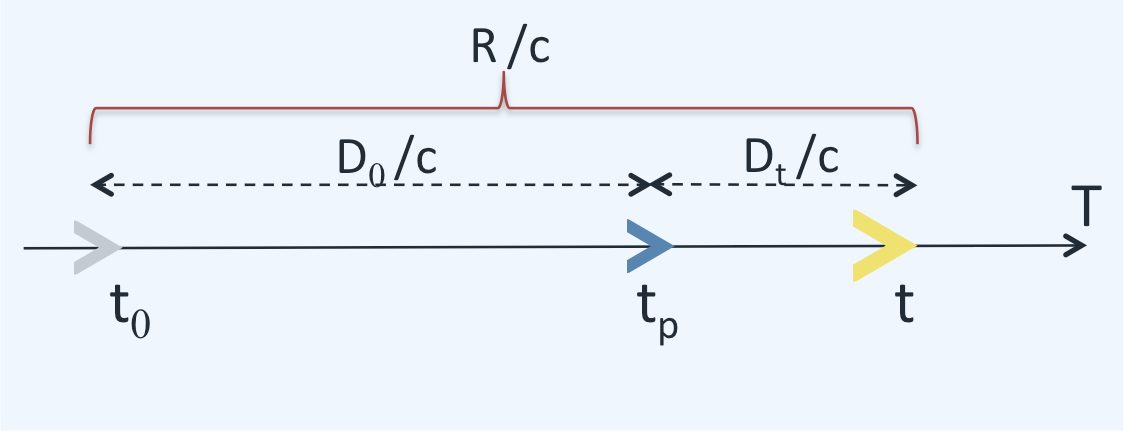}
\caption{
The figure shows the three time parameters $t_0$, $t_p$ and $t$ as defined in section \ref{spacetime}, coherently to the example of fig.\ref{spacex}.
\label{timet} }
\end{figure}

The previous quadratic equation can be solved obtaining two terms 
\be
t_0=\frac{t_p-\sqrt{\beta_0^2t_p^2+y_0^2(1-\beta_0^2)/c^2}}{1-\beta_0^2}
\label{tt1}
\ee
and
\be
t=\frac{t_p+\sqrt{\beta_0^2t_p^2+y_0^2(1-\beta_0^2)/c^2}}{1-\beta_0^2}.
\label{tt0}
\ee
Respectively, the two solutions $t_0$ and $t$ represent the time birth of the signal and  time observation at $[0,y_0]$ point. So, their difference corresponds to the interval $\Delta t$ as defined by eq.\ref{rt1}. The source corresponding positions $x_0$ and $x_t$ are illustrated in fig.\ref{spacex}. They represent also the analytic solutions of the quadratic eq.\ref{xpxt}. So they can be written in the following way
\be
x_{0,t}=\frac{x_p\pm \beta_0 \sqrt{x_p^2+y_0^2(1-\beta_0^2)}}{1-\beta_0^2}
\ee
Placing $D_0\equiv \sqrt{x_0^2+y_0^2}$ and performing some algebra, the following formulas are obtained
\be
R=D_t+D_0=
\label{r_sum}
\ee
\be
=2\gamma_0^2(D_t-\beta_0 x_t)=
\label{rt2}
\ee
\be
=2\gamma_0\sqrt{x_p^2\gamma_0^2+y_0^2}
\label{xpi}
\ee
where $\gamma_0\equiv 1/\sqrt{1-\beta_0^2}$.\\
The terms $x_t$ and $x_0$ are also related by
\be
x_t-x_0=v_0(t-t_0)=\beta_0 R.
\label{x0}
\ee
The space quantities defined in the present section are illustrated in the fig.\ref{spacex}, while the three time parameters are shown in the fig.\ref{timet}.
In the expression of R, in particular in \ref{xpi}, it is easy to recognize the same result as  Feynman \cite{fey}, obtained referring it to the $retarded~potential$ of Li\'{e}nard-Wiechert. In the same reference, how formalism is directly derived from Maxwell's laws is shown, and the author also reaffirms that the same procedure was used to get to the Lorentz transformations (LTs). Although expressing the same formalism, a direct comparison between the LTs and the relationships from the present calculation is difficult. The problem is that Lorentz considers two independent instants ($t$ and $t'$), while in this treatments there are different time values but they refer to same physical state recorded at instant $t$. They are $t_0$, the time origin of $R$, $t$ the signal recording time, and $t_p$, an intermediate instant representing the time origin of $D_t$.  Although $t_0$ and $t$ are linked by a linear relationship as defined by \ref{rt1}, this is not true among $t_p$ and the other two values, as it shown by the eqs.\ref{tt1} and \ref{tt0}. Accordingly, confusing the time values or their mutual roles would alter the geometric rates and even predictions about the system dynamics. To avoid such ambiguities in the laws of physics, then we must consider the amount $ R $ instead of the simple distance $D_t$, since the first one is generated by a time linear relationship \ref{rt1}, while $ D_t $ is defined by the quadratic eq.\ref{xp1}.
\begin{figure}
\includegraphics[width=0.9\linewidth]{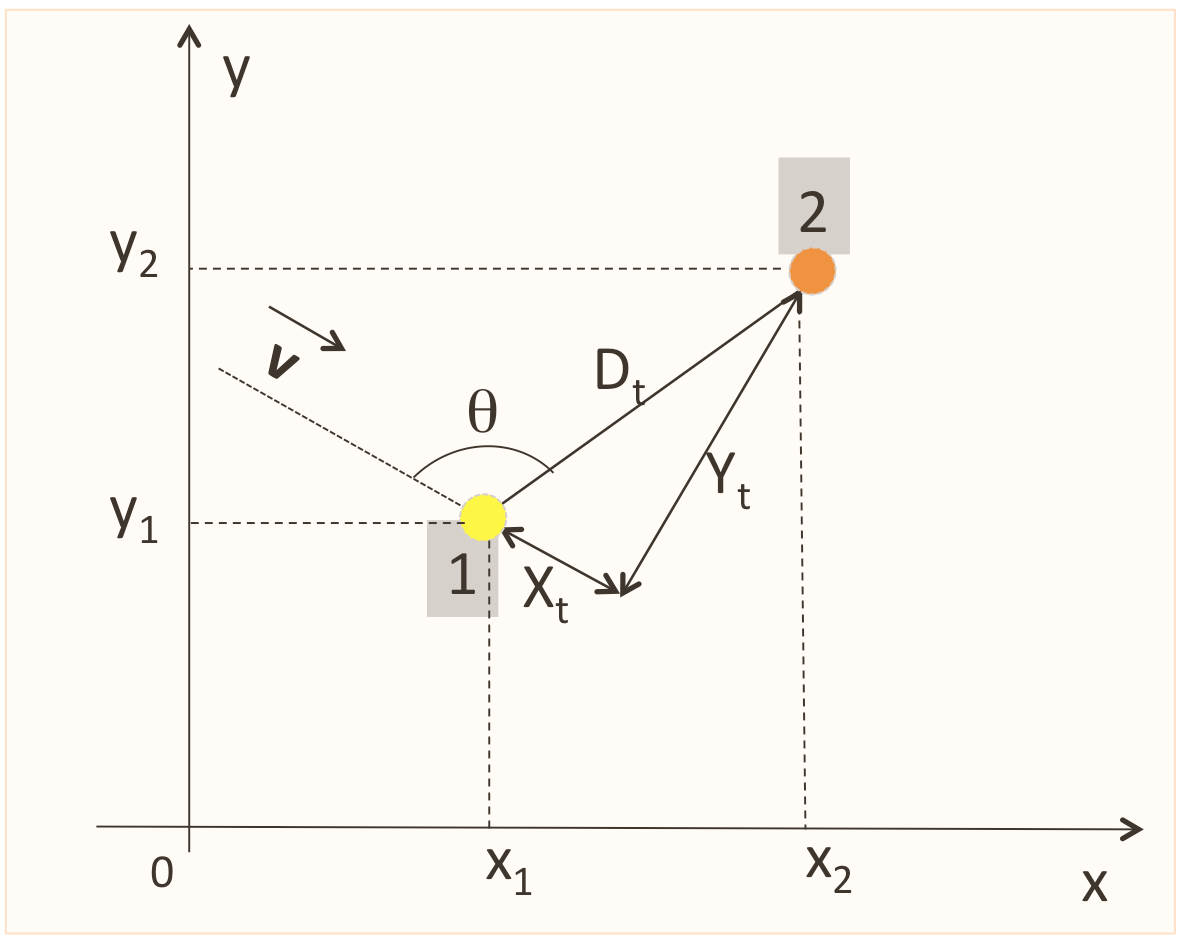}
\caption{
Emitting a signal at $c$ speed, a source travels at velocity $\bf v$. At time $t$ the distance between observer$\bf 1$ and source$\bf 2$, is $D_t$. The lengths $D_t$ and $X_t$ are invariant with respect to any reference system, as well as the corresponding value of $R$ (eq.\ref{rnew}), thus respecting the Galilean invariance.
\label{refe} }
\end{figure}
\section{invariance and generalization of the reference system}
Let's consider a reference system $S(x,y)$ as in the fig.\ref{refe}, where the source has coordinates $(x_1,y_1)$ and the observer $(x_2,y_2)$. In a general case, before interaction or simple observation, both the source as the detector can move at $\bf v_1$ and $\bf v_2$, each one following its own physical law. The difference between the velocities
\be
{\bf v}\equiv{\bf v_1}-{\bf v_2}
\label{vti}
\ee
then represents the relative velocity between the objects belonging to the non-perturbed system.
Due to movement of both points $\bf 1$ and $\bf 2$, the distance $D_t$ will vary following a double time law. To express that, it is useful to define the couple $X_t,Y_t$ both depending on time, such as
\be
%X_t\equiv\frac{|x_2-x_1|}{x_2-x_1}\sqrt{D_t^2-Y_t^2},
X_t\equiv D_t\cos\theta~~~{\text{and}}~~~Y_t\equiv D_t\sin\theta
\label{xnew}
\ee
where $\theta $ is the angle between ${\bf v}$ and $D_t$, so
\be
D_t=\sqrt {X_t^2+Y_t^2}
\label{dtnew}
\ee
Despite the quantities $X_t$ and $Y_t$ obey to an analytical geometry law, the locus  $(X_t,Y_t)$ can not be assumed as cartesian coordinate system, given that its origin would vary instant by instant. Rather, they identify a virtual space $\Phi (X,Y)$ where the following relationships are valid
\be
R(t_0,t)\equiv c(t-t_0)=2\gamma^2(D_t-\beta X_t)
\label{rnew}
\ee
with
\be
\beta\equiv\frac{|{\bf v}|}{c}~~~{\text{and}}~~~\gamma\equiv 1/\sqrt{1-\beta^2}.
\label{betanew}
\ee
As such, the expression for $R$ can be rewritten in any reference frame while maintaining the relationship between the basic parameters $D_t$, $X_t $ and $Y_t$, and thus respecting the Galilean invariance.
\section{ examples of space-time laws}
Determining $D_t$, the degrees of freedom are many as are the dimensions of space, so three in general. Each one follows its temporal law. In this representation we have just two, $X_t$ and $Y_t$, having chosen a two-dimensional space (with z = 0). For the complete resolution of the eq.\ref{rnew} and so to define $R$, it is also necessary to know $\bf v$ which generally depends on the initial conditions. Following, some typical field configurations.
\subsection{Electromagnetic signal observation}
\begin{figure}
\includegraphics[width=.9\linewidth]{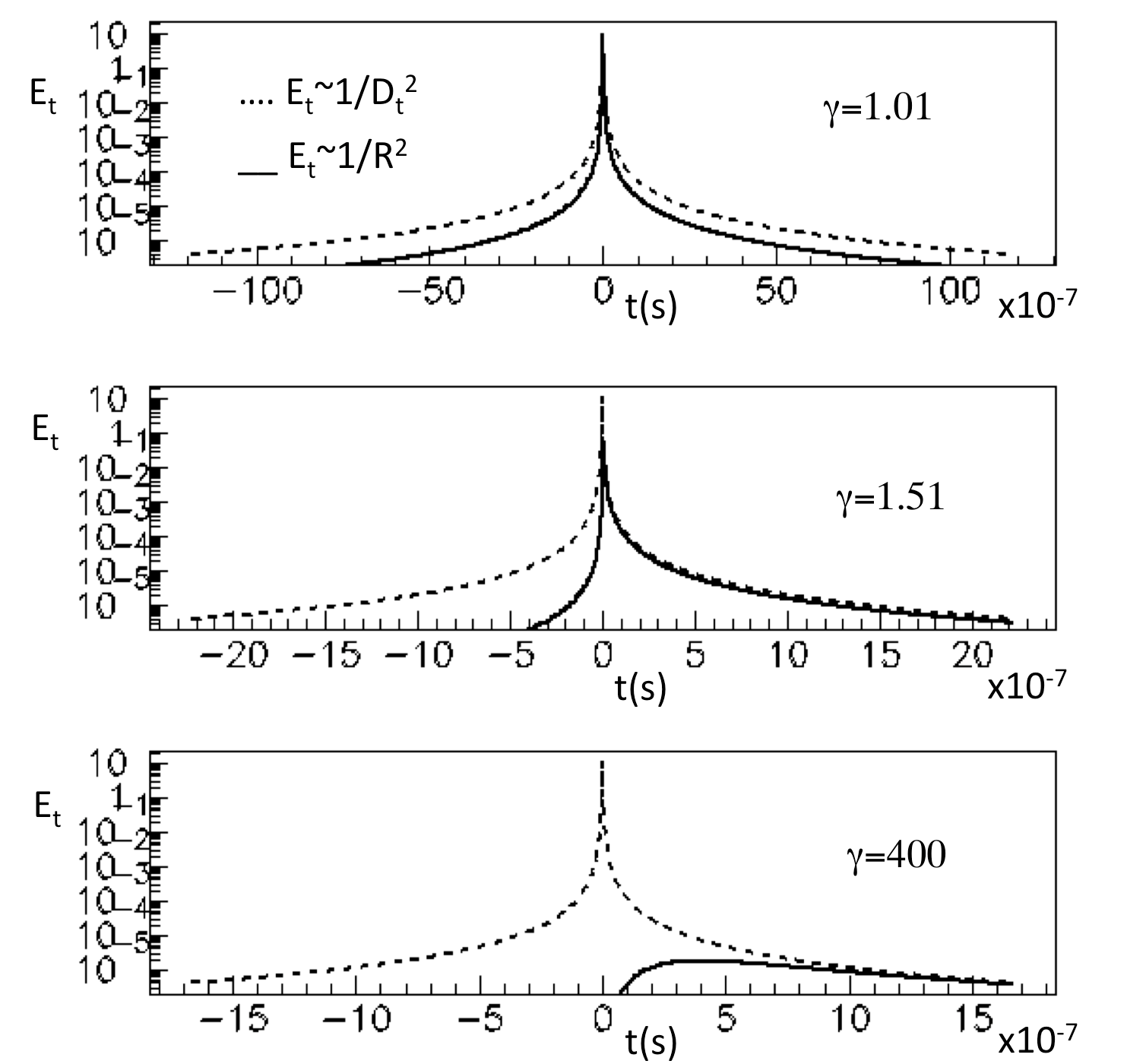}
\caption{
The time patterns of field intensity $E\sim1/R^2$ (solid line)  are shown for three values of $\gamma_0$, from top respectively $1.01,~1.51,~ 400$. The dotted line is the prevision based on $E\sim1/D_t^2$.  The difference between them is evident to growing of the $\gamma$ parameter. The detector position is fixed at $[0, 30cm]$.
\label{gamma101} }
\end{figure}

\begin{figure}
\includegraphics[width=0.9\linewidth]{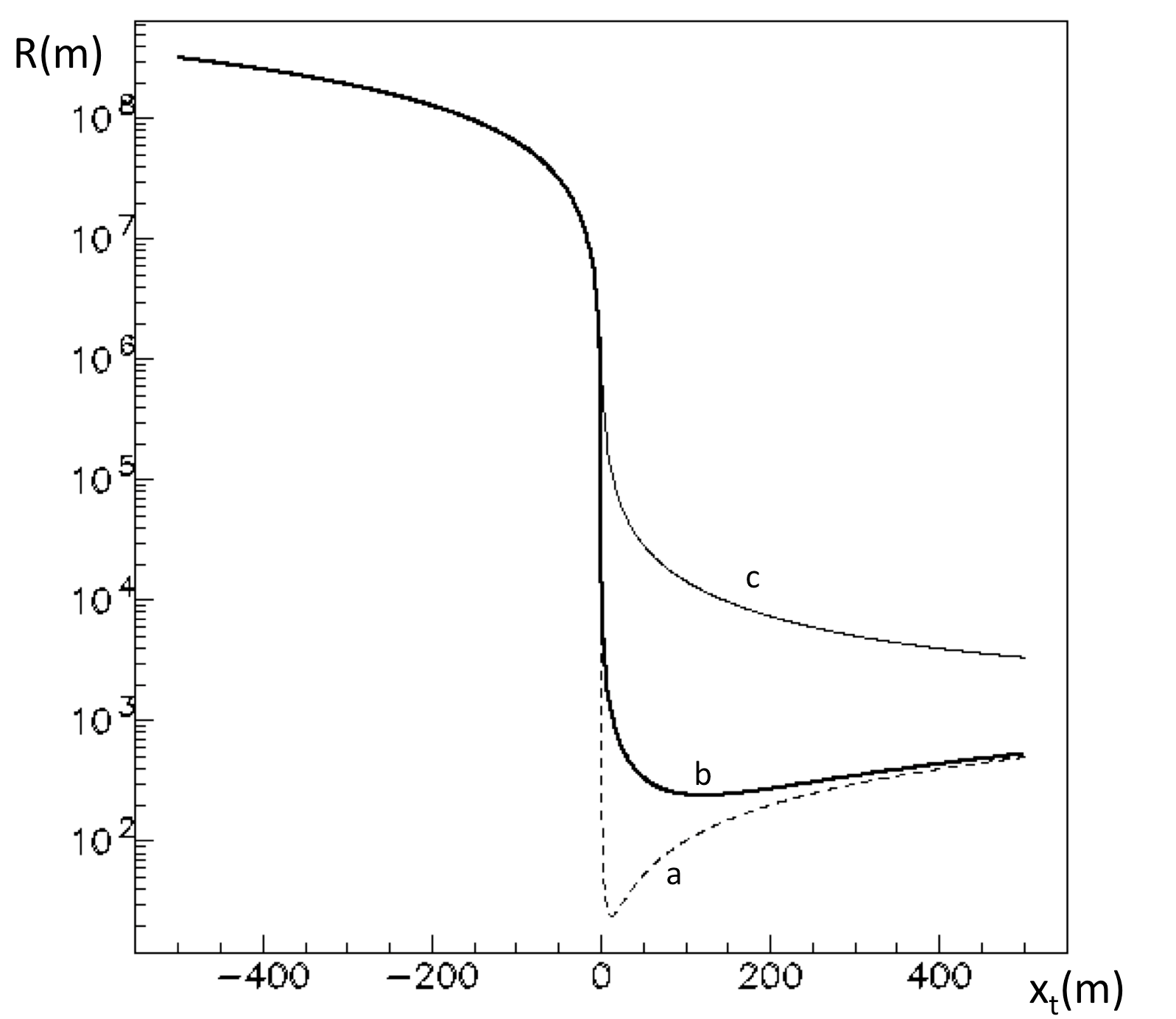}
\caption{
The figure refers to the case $\gamma=400$,  but with three different lengths for $[0,y]$, the detector ordinates. The levels of $R$, $a$ $b$ and $c$,  correspond to $3cm$, $30cm$ as the third  part of the previous example, and $3m$. Non-linear effects are impressive especially changing the $x_t$ sign.  
\label{caso3x3} }
\end{figure}

In the section \ref{example1}, observing the signal from a moving source, a linear law had chosen for $X_t$, and a constant term for $Y_t$. 
Indeed, being $v_0$  constant in magnitude and direction, and remembering that by definition $X_t$ is parallel to $ \bf v$, the choice of the reference system
\be
S(x,y)\equiv \Phi (X,Y)
\label{ref2}
\ee
 is not only possible but even natural. Thus
\be
X_t=x_t=v_0t~~~{\text{and}}~~~Y_t=y_0=constant.
\label{exe1}
\ee
Assuming this configuration and the eq.\ref{rnew} (or equivalently eq.\ref{rt1}), we outline the intensity field, with a few numerical example.
In fig.\ref{gamma101}, considering both $E\sim 1/R^2$ and $E\sim 1/D_t^2$, the time behaviors are shown for three values of $\gamma_0$, $1.01,~1.51,~ 400$. The example refers to the case with $v_0=\approx 0.6c$, considering $c$ as light speed\cite{data}, and being the detector fixed to $y=0.3m$. It is evident the difference between the two predictions, as to $\gamma_0$ grows. In our opinion, $E\sim 1/R^2$ is right, thus confirming the significance of $t$ as present time and the significance of $R$ as the real signal length.
For $\gamma_0=400$, the behavior of $R$ is calculated considering three values for the detector distances $[0,y_0] $. The fig.\ref{caso3x3} shows how this geometry parameter is physically critical, since some difference not only can generate distant signals many orders of magnitude between them, but also morphologically very different.
\subsection{Stationary orbit}
\begin{figure}
\includegraphics[width=.95\linewidth]{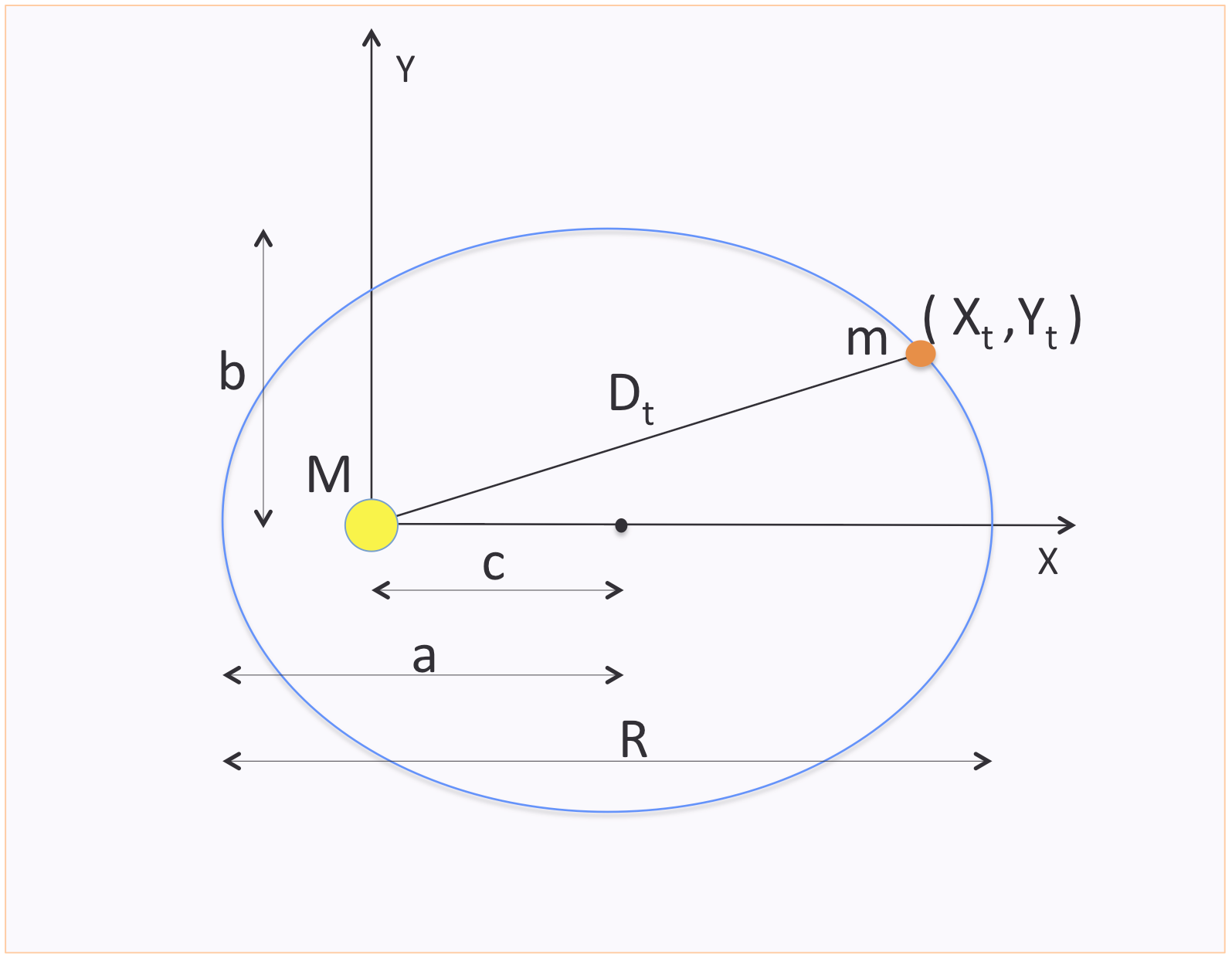}
\caption{
Equipotential orbit described by the object with mass $m$ around $M$. It is the set of $(X_t,Y_t)$ points solving the eq.\ref{rnew} with constant $R$, and it coincides with an elliptical orbit with specific parameters as indicated by the statements \ref{elli1}-\ref{epsi}.
\label{fig_elli} }
\end{figure}

\begin{figure}
\includegraphics[width=.90\linewidth]{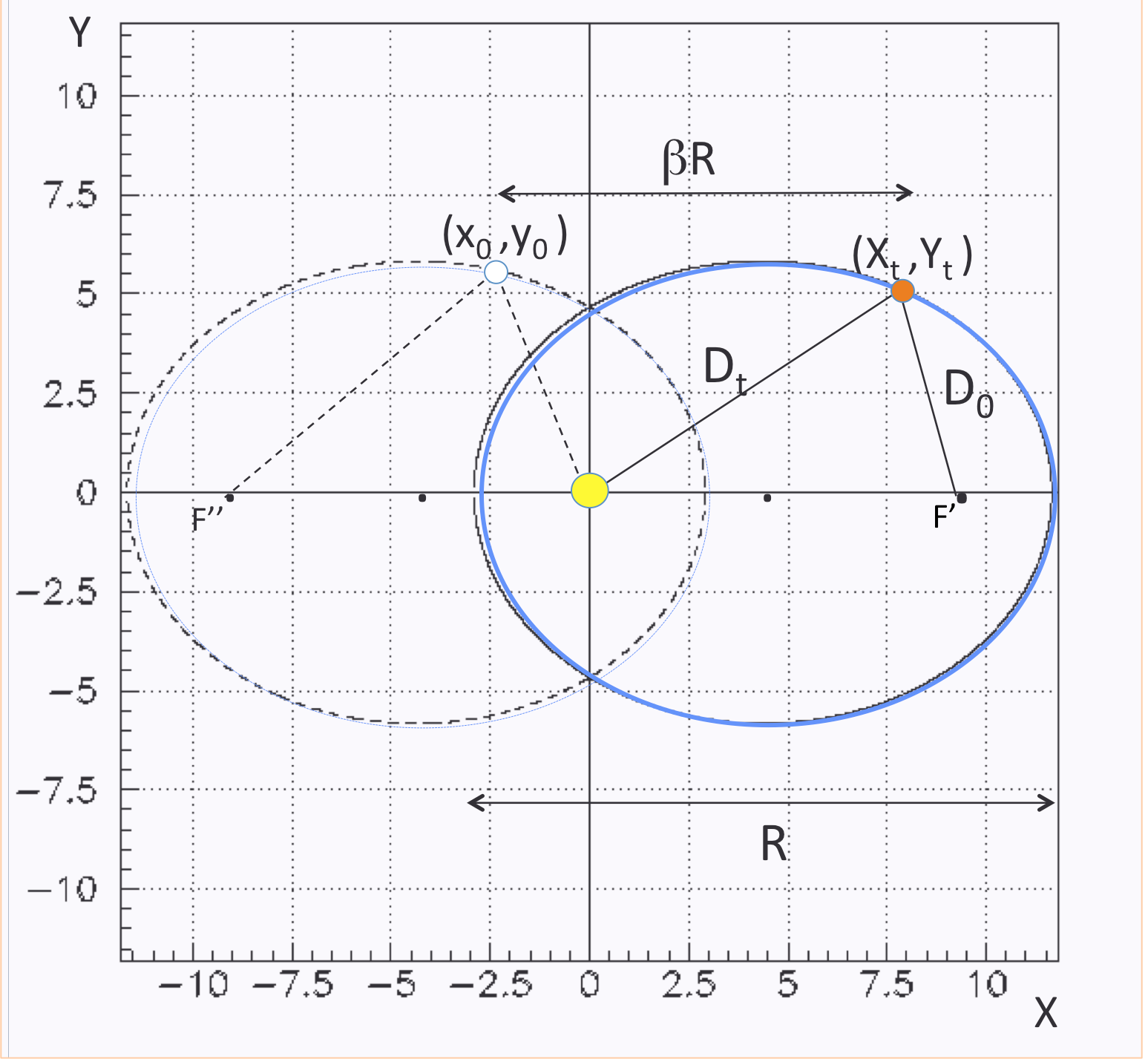}
\caption{
The same phenomenology of fig.\ref{fig_elli} is shown adding the locus $(X_0,Y_0)$, dotted elliptical line on the left, solving $X_0=X_t-\beta R$, and $R=2\gamma^2 (D_0+\beta X_0)$.  
\label{fig_elli2} }
\end{figure}

The present study can be applied to electromagnetic case as well as to a gravitational source, since  both cases follow the inverse-square-law, as described by the relation \ref{et1}. In particular, we presume a body of mass $m$, running around a gravitational source with mass $M>>m$. According to the law \ref{et1}, there is a potential energy difference $\Delta U\sim 1/R$ between the two masses. 
Assuming that the two objects are moving keeping constant the potential energy difference, we obtain
\be
\Delta U\sim 1/R=constant \Rightarrow R=constant.
\label{deltau}
\ee
Initially, before interacting, $M$ and $m$ moved at $\bf v_M$ and $\bf v_m$ whose difference was $\bf v$.
Keeping constant $R$ and $\beta=|{\bf v}|/c$, it is possible verify that in the reference system $\Phi(X,Y)$, the locus ($X_t,Y_t$) which satisfies the eq.\ref{rnew}, is an ellipse with the geometric parameters $a$, $b$ and $c$, respectively major and minor semiaxis, and half distance between the two foci, equal to
\be
a=R/2~~~~b=R/2\gamma~~~~c=\beta R/2,
\label{elli1}
\ee
 as shown in figs.\ref{fig_elli} and \ref{fig_elli2}.
One of two foci is centered into the gravity center, or equivalently in the origin of coordinate system $\Phi(X,Y)$,
\be
F_1\equiv (0,0),
\label{focus}
\ee
further, the ellipse eccentricity $\epsilon$ results
\be
\epsilon=\beta.
\label{epsi}
\ee
These statements confirm the  Kepler's first law, and also reveal a strong analogy between geometrical and physical properties in such dynamics.
\subsection{Local Solar System}
To test the effectiveness of eq.\ref{rnew} and \ref{elli1}-\ref{epsi}, we apply them to a real case, for instance, to the local solar system constituted by the Sun and eight planets plus Pluto. We consider the orbits of the planets as extremely conservative systems and then we set $R$ and $\bf v$ constant in the resolution of eq.\ref{rnew}, being $\bf v$ the difference between $\bf v_M$ the solar speed, and $\bf v_m$ the free motion of the planet. Reading the actual data from the standard tables \cite{nasa}, essentially extracting $a$ and $b$, lengths of major and minor semi axes of the planets orbiting in the solar system, from that the relative eccentricity $\epsilon$ values, we can obtain the relative values of $R$ and $\beta$, using the equivalences $R=2a$ and $\beta=\epsilon$. 
\begin{figure}
\includegraphics[width=0.85\linewidth]{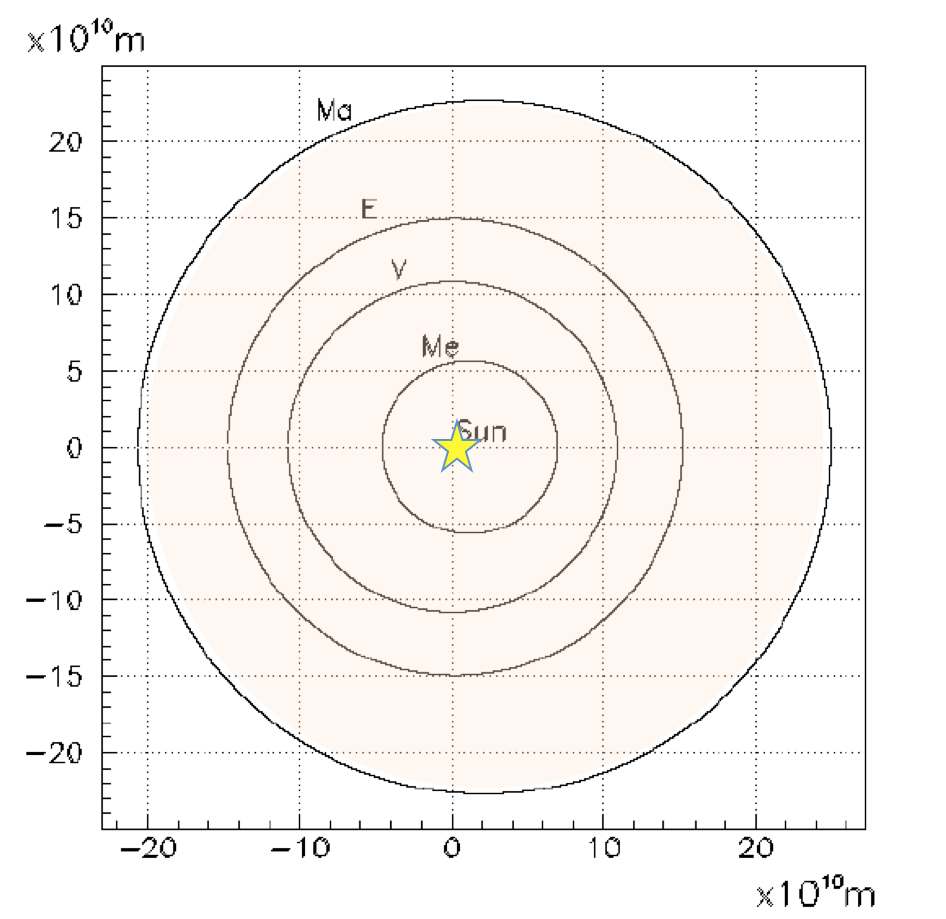}
\caption{The trajectories $(X_t,Y_t)$ of the four planets nearest the Sun are drawn. They have been calculated by eq.\ref{rnew}, taking into the account the  equivalences \ref{elli1}-\ref{epsi} and the eq.\ref{rnew}. Sun place coincides naturally with the origin of the axes intersection. 
\label{sun1} }
\end{figure}
\subsubsection{Orbits}
\begin{figure}
\includegraphics[width=0.95\linewidth]{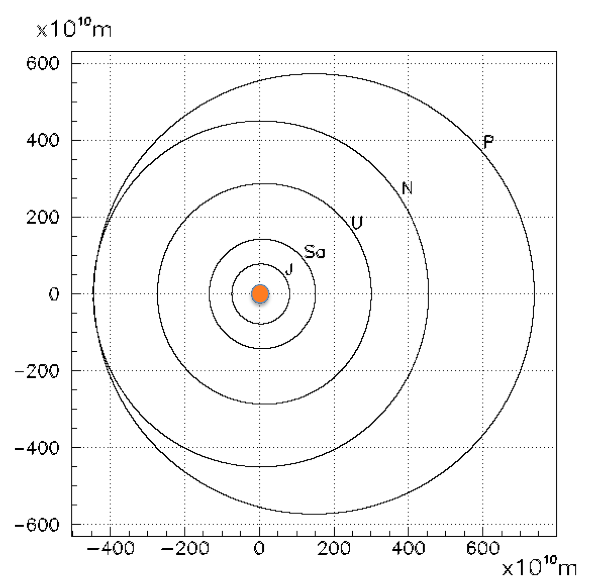}
\caption{The orbits of the four planets more distant from Sun are shown, plus that of Pluto, the largest one. As in the previous figure, they derive from the statements \ref{elli1}-\ref{epsi}, and solving the eq.\ref{rnew}.
\label{sun2} }
\end{figure}

Being $D_t=\sqrt{X_t^2+Y_t^2}$, the couple $(X_t,Y_t)$ solving the eq.\ref{rnew} coincides with the actual orbits described by the nine objects, as in the fig.\ref{sun1} and fig.\ref{sun2}. In this case, the real space $S(x,y)$ and the virtual one $\Phi(X,Y)$ can be considered almost coinciding probably due to the fact that $\bf v$, the difference between the solar speed $\bf v_M$ and $\bf v_m$, is kept constant in the dynamics of this system. So, the planet orbits naturally describe an equipotential path, and are really elliptical, with the Sun, the center of mass, placed  at one of the two foci, as provided by the Kepler's first law.
\subsubsection{Orbital speed}
The orbital speed is evaluated for each of the nine bodies belonging to the solar system, assuming the Sun\cite{data} as center of mass. Coming after some logical steps, the following relation can be adopted for this as well as for any system with center of mass $M$
 \be
 v_{orb}=\sqrt{\frac{2\gamma^2 GM}{R}\left[1+\beta^2-\frac{2\beta X_t}{D_t}\right]}.
 \label{v_orbi2}
 \ee
 where $G$ is the gravitational constant.
The eq.\ref{rnew} is applied along the whole trajectories, tabulating the average with the minimum and maximum values in tab.\ref{tab1}. 
 \begin{table}[]
%\centering
\caption{Velocity values and orbital period for the planets of the solar system, including Pluto. The formula \ref{v_orbi2} is applied. }
\label{tab1}
\begin{tabular}{c|clcl}
%& & & &\\
 &$  v_{(aphelion)}$   & $~~~\bar{v}$& $v_{(perihelion)}$ \\
 &$  (Km/s)$   & $~(Km/s)$& $~(Km/s)$ \\
 \hline
 Mercury & 38.858 &48.215  & 58.976 \\
 Venus& 34.784 & 35.020 &  35.258 \\
 Earth& 29.291 & 29.786 &30.286    \\
 Mars& 21.972 & 24.164 & 26.497  \\
 Jupiter& 12.440 & 13.063 & 13.705  \\
 Saturn&9.138  &9.649 & 10.179 \\
 Uranus& 6.485 &  6.802& 7.128  \\
 Neptune& 5.385 &5.432  & 5.478  \\
 Pluto& 3.676 & 4.790 & 6.112  \\
 \hline
\end{tabular}
\end{table}
These two values correspond to the object positions relatively at aphelion and perihelion, as expected from Kepler's second law. The agreement with the measurements \cite{nasa} is good especially for the minimum and maximum values that regard punctual body positions. Some small discrepancies about the average value can be due to variability of density of states along the each orbit, and to the different algorithms used for the elaboration of the average value.
\section{conclusion}
Based on euclidean space-time geometry and on analytic procedures, the quantity $R$ is revealed. It doesn't  correspond simply to geometric distance between source and detector, rather it is the real signal path  covered at $c$ speed. It  represents the physics magnitude to take into the account for the  intensity field evaluation, in any $inverse~square~law$. In light of this new quantity, classical physical paradigma have been re-examined, bringing out unsuspected as natural analogies between the geometric and physical parameters describing the mechanics of moving bodies. Testing the method on a newtonian system such as the local solar system, the result perfectly fits to the actual data. The assessment  is compatible with relativistic formulas although no hypothesis was formulated on light special isotropy, neither on the space-time curvature, posing the issue on the actual need to introduce these concepts in interpreting the dynamics of moving bodies. \\

{\large\bf Acknowledgements }\\

I would like to thank my precious teachers and colleagues who inspired this study, even the many who have allowed to achieve it, with their helpful suggestions and unfailing encouragements. 
I am also grateful to Giorgio Fornetti for his valuable advices and his thorough review of the manuscript.
%does not object to the pursuit of truth.


\begin{thebibliography}{99}
\bibitem{luce}  Generally, $c$ indicates the speed of light but from the analytical point of view, it represents simply the emission velocity of any signal, without other special physical hypotheses.
\end{thebibliography}
 \end{document}